# Intrinsic polarization conversion and avoided-mode crossing in X-cut lithium niobate microrings

*Zelin Tan, Jianfa Zhang, Zhihong Zhu, Wei Chen, Zhengzheng Shao\*, Ken Liu\*, and Shiqiao Qin\**

Zelin Tan, Jianfa Zhang, Zhihong Zhu, Ken Liu, and Shiqiao Qin

College of Advanced Interdisciplinary Studies & Hunan Provincial Key Laboratory of Novel Nano Optoelectronic Information Materials and Devices, National University of Defense Technology, Changsha, 410073, China.

Nanhu Laser Laboratory, National University of Defense Technology, Changsha, 410073, China.

E-mails: liukener@163.com, sqqin8@nudt.edu.cn

Wei Chen

College of Meteorology and Oceanography, National University of Defense Technology, Changsha, 410073, China.

Zhengzheng Shao

School of physics, Central South University, Changsha, 410083, China

E-mails: zzshao_nudt@163.com

Abstract: Compared with well-developed free space polarization converters, polarization conversion between TE and TM modes in waveguide is generally considered to be caused by shape birefringence, like curvature, morphology of waveguide cross section and scattering. Here, we reveal a hidden polarization conversion mechanism in X-cut lithium niobate microrings, that is the conversion can be implemented by birefringence of waveguides, which will also introduce an unavoidable avoided-mode crossing. In the experiment, we find that this mode crossing results in severe suppression of one sideband in local nondegenerate four-wave mixing and disrupts the cascaded four-wave mixing on this side. Simultaneously, we proposed, for the first time to our best knowledge, one two-dimensional method to simulate the eigenmodes (TE and TM) in X-cut microrings, which avoids the obstacle from large computational effort in three-dimensional anisotropic microrings simulation, and the mode crossing point. This work will provide an entirely novel approach to the design of polarization converters and simulation for monolithic photonics integrated circuits, and may be helpful to the studies of missed temporal dissipative soliton formation in X-cut lithium niobate rings.

## 1. Introduction

As one of the basic properties of electromagnetic waves, polarization plays important roles in optical communication and detection [1]. Leveraging birefringence of structures, originating from shape or material itself, is the main method to manipulate the polarization, which has been developed well in free space like converters based on metasurfaces [1-4]. However, it remains a challenge to interconnect this kind of converters with other integrated functional devices, which can be handled by fiber [5], but greatly destroy the miniaturization of integrated circuits. In waveguide regime, polarization conversion of orthogonal modes, TE and TM, is generally regarded as mode coupling, arising from the curvature, cross-section size, sidewall inclination angle and scattering of the waveguide [6-10], which can be classified as shape anisotropy. Several papers have demonstrated that polarization coupling or mode hybridization in thin-film lithium niobate (LN) waveguides, however, the essence is not figured out clearly [11-14]. As a result, the potential of birefringent platform, such as lithium niobate (LN), for mode conversion has not been excavated.

In recent years, with the maturity of commercial manufacture of thin-film LN and breakthroughs in low-loss LN waveguide fabrication [15-19], lithium niobate on insulator (LNOI) has become one of the most attractive platforms for monolithic PIC, due to its excellent nonlinear, acousto-optic, electro-optic characteristics, relatively high refractive index and wide-spectrum transparent window [20]. In practices, to make full use of the excellent properties of LN, such as the second-order nonlinear tensor element $d_{33}$(-27 pm/V) and Pockels coefficient $r_{33}$ (31 pm/V), devices are usually designed and fabricated on X-cut LN platform. Common structures include Mach–Zehnder interferometer-based modulators capable of ultrafast modulation [21-23], periodically polarized LN waveguides for wavelength conversion [24, 25], and rings for generating optical frequency combs, etc. [26, 27].

Here, by investigating the nondegenerate four-wave mixing (FWM) in a X-cut LN microring, where the birefringence cannot be neglected or avoided, we find that one sideband (blue wavelength) of a pair is suppressed severely with two pumps detuned from blue to effective zero detuning of their corresponding high-Q resonances. The suppressed sideband is released until the third sideband (red wavelength) is generated by cascade FWM. From the resonator transmission spectrum, we find that two mode families (TE and TM) are supported, resulting in avoided-mode crossing between the pump and first sideband in blue wavelength. Thus, the non-degenerate FWM is disrupted as well as subsequent cascade FWM in this side for heavily distorted dispersion around the crossing position.

The reason for this phenomenon is a hidden mechanism in birefringent waveguides. Because the angle between TE mode polarization and optical axis (z axis) changes with light propagation in the ring, the effective index of the TE mode varies with azimuth on the interval $[n_e, n_o]$, where $n_e, n_o$ represent the extraordinary and ordinary index. For TM mode, its polarization is always perpendicular to the z axis, so the effective index corresponds to ordinary index abidingly. This distinction causes the effective indices of these two modes in X-cut LN microrings will cross at certain azimuth, so that modes show strong interaction at these points resulting in polarization conversion, stimulating the TM modes when satisfy the resonance condition. This is inevitable and will disrupt the FWM around the mode-crossing point between TE and TM mode families. Calculating the eigenmodes as well as mode crossings point in a birefringent microring is not easy, which has not been reported to the best of our knowledge, due to the anisotropic nature and the huge computational effort. At the end of this paper, we propose a 2D-equivalent method, stretching the ring to a straight waveguide with periodic boundaries at both ends, that fits well with experimental results without considering the dispersion.

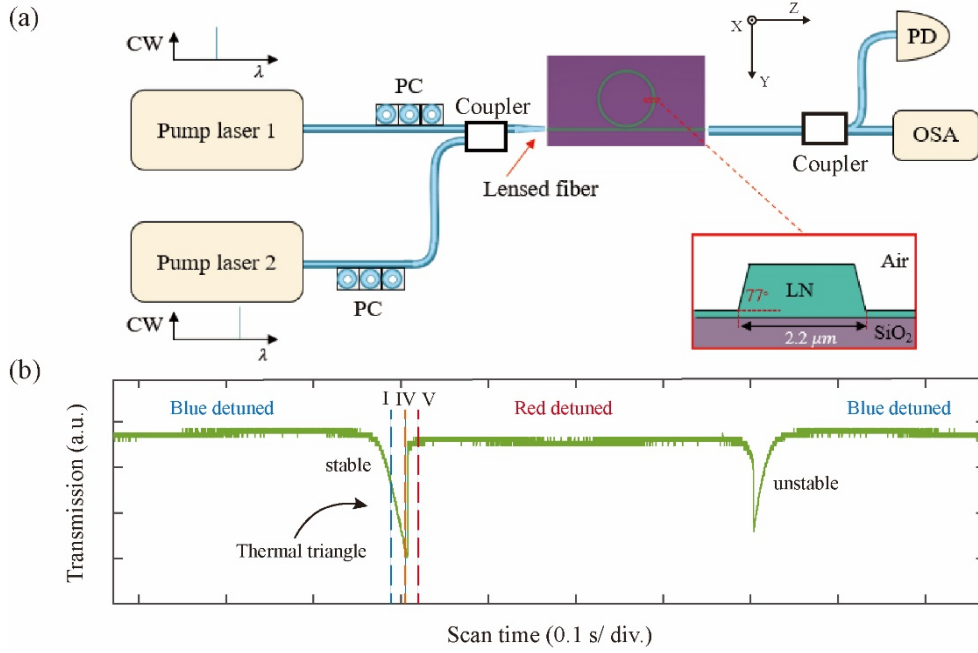

**Figure 1.** a) Experimental setup. PC: Polarization controller; PD: Photodetector; OSA, Optical spectrum analyzer. The coordinate system in the figure corresponds to the actual crystallographic. Inset: schematic of waveguide cross section. b) Forward and backward laser-scanned transmission spectrum near 1602 nm, under 4 mW pumping. The effective resonance is distorted as a triangular shape called "thermal triangular" due to thermal absorption. The dashed lines I, IV and V in the figure indicate the three relative positions of pump in the resonance respectively.

## 2. Results and Discussion

**Figure 1a** shows the experimental setup for nondegenerate FWM. Two continuous wave (CW) lasers with different wavelengths (~1602 nm, ~1594 nm) and same power (~4 mW) are coupled into the waveguide through polarization controller and fiber coupler successively. The inset in Figure 1a shows a schematic cross-section of the waveguide. In this experiment, an X-cut LN microring is fabricated by $Ar^+$ ion. A total of 615 nm of LN was etched, leaving a 75 nm thick slab. The waveguide sidewall tilt angle is 77°. The ring radius is 23 um, corresponding to a Free Spectral Range (FSR) of ~1 THz. When CW pump is coupled into the cavity modes, the cavity absorbs a small fraction of the energy and heated up. Elevated temperature will change the effective refractive index of the cavity mode, shifting the resonance frequency in the scan direction towards longer wavelength. One kind of thermal feedback is established as resonance locked to the pump laser passively. As a result, the Lorentz-shaped adiabatic resonance is distorted to a "thermal triangular" [28], as shown in Figure 1b. Because of this thermal effect, the pump can only be tuned gradually from short wavelength to intracavity mode. Conversely, if pump is tuned from red detuned, the coupling to resonance is unstable. The dashed lines I, IV and V in Figure 1b indicate the positions of pump at different stages relative to intracavity mode during the pump tuned from blue detuning to intracavity mode for FWM.

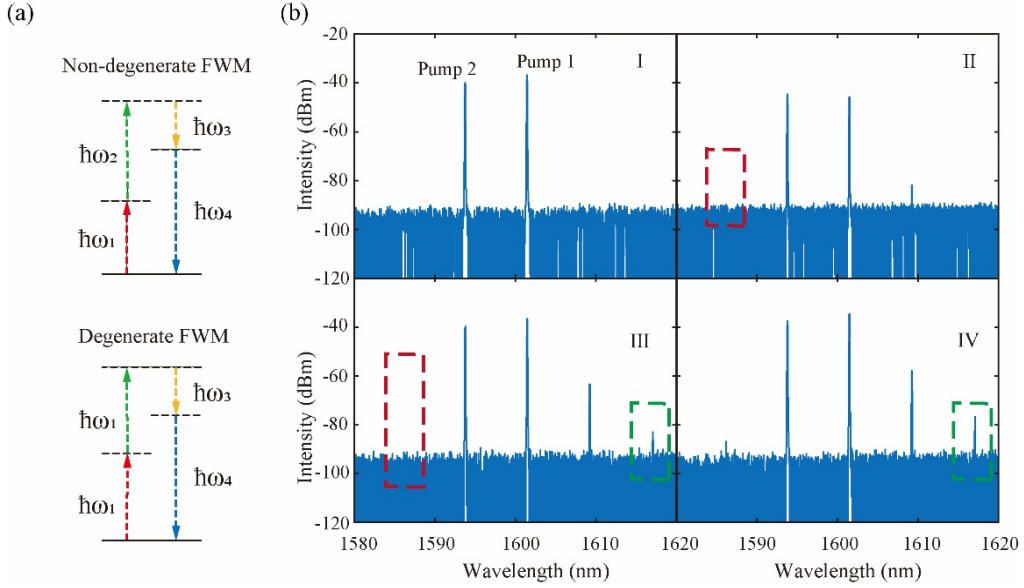

**Figure 2.** a) Illustration of annihilation (upwards pointing arrows) and creation (downward pointing arrows) of photons in nondegenerate and degenerate four-wave mixing processes. The length of the arrows corresponds to the energy of photons. b) Spectra of four stages in FWM with two pumps (Pump 1 and Pump 2) which correspond to two adjacent intracavity modes, I-IV, during two CW pumps tuning from blue detuning to effective zero detuning of resonances. The red dashed boxes in stages II and III indicate the position of suppressed sideband in the FWM process. The rightmost sideband (inside green dashed boxes) in stages III and IV is generated by cascaded FWM. Stage IV indicates that when the two pumps are completely coupled into the intracavity modes.

FWM can be divided into nondegenerate and degenerate process, as shown in **Figure 2a**. The corresponding nonlinear polarization are:

$$P_{NL}^{non} = 6\,\epsilon_0\chi^{(3)}E_1E_2E_3^* e^{i(k_1+k_2-k_3)z}e^{-i(\omega_1+\omega_2-\omega_3)t}, \tag{1}$$

$$P_{NL}^{de} = 3\,\epsilon_0\chi^{(3)}E_1^2 E_3^* e^{i(2k_1-k_3)z}e^{-i(2\omega_1-\omega_3)t}. \tag{2}$$

To achieve maximum conversion efficiency, energy conservation between pump and signal should be satisfied, as well as phase matching, like $k_1 + k_2 - k_3 = k_4$ for nondegenerate four-wave mixing. In the experiment, two pumps are coupled into two adjacent intracavity modes assisted by a photodetector slowly from blue detuning, for self-locking into the "thermal triangular" induced by thermal effect. Meanwhile, the other pump-induced change in refractive index, affects the detuning of the respective parties. After the pumps are all roughly tuned into the resonances, as shown by dashed line I in Figure 1b, there is no sideband generated since the intracavity energy is low and the phase is remarkably mismatched. With two pumps (1, 2) tuned to long wavelength gradually, the first sideband in red wavelength is obtained, as shown in stage II in Figure 2b, while there is supposed to be one sideband on the symmetrical side (red dashed box) based on the two conservations. The sideband intensity is not high, due to weak interaction between pumps and cavity modes. Continue to tune the pumps to slight detuning, and a new sideband (in blue box), generated by cascaded FWM on the red side, as shown in stage III of Figure 2b, while the suppressed sideband on blue side is still missing. Stage IV depicts the spectrum when both pumps are in zero detuning to the resonances, as the dashed line IV in Figure 1b. At this moment, the alterations of refractive index caused by thermal and photorefractive effect in LN ring tends to be stable, pumps can be fully coupled into the cavity mode stably, the energy in the cavity is high and the phase is matched. So, the suppressed sideband is released besides the intensity of sidebands generated in previous stages enhanced. Further tuning to red of pumps will break down the self-locked dynamic, so that no energy is coupled into the cavity instantly and all sidebands are eliminated, as the dashed line V in Figure 1b.

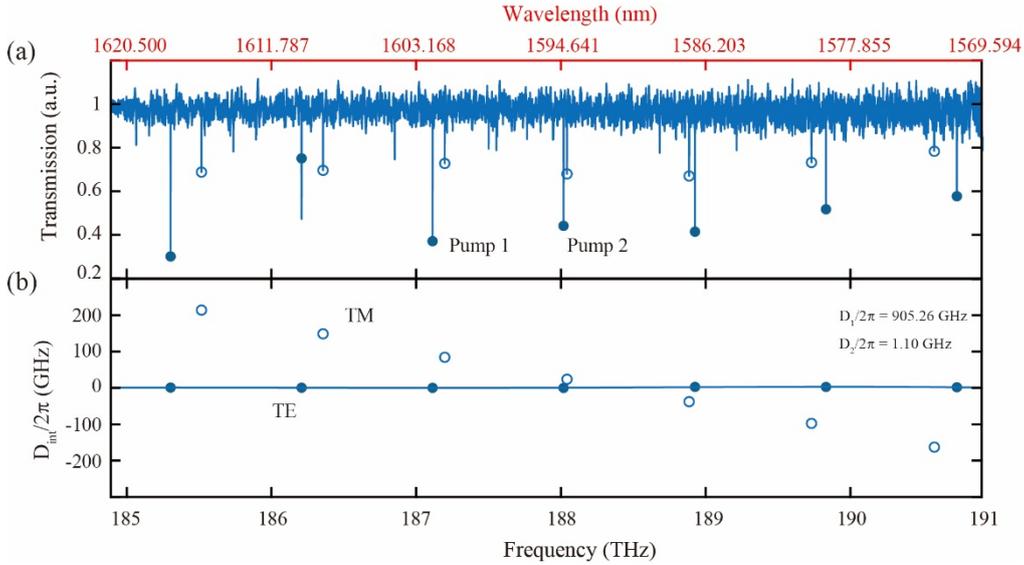

**Figure 3.** a) Normalized cavity transmission spectrum with solid circles denoting TE mode and hollow circles denoting TM mode. b) The dispersion $D_{int}/2\pi$ of TE mode. The solid circles correspond to the TE mode in (a), and the blue line is a fourth-order dispersion fitting.

Generally speaking, non-degenerate FWM is easy to be implemented because of the small interval of each component, which can be regarded as zero dispersion. However, for 1-THz-FSR rings, the cavity mode spacing is too large to smash the effect of group velocity dispersion. The presence of dispersion makes the actual resonances not exactly equidistant, and their angular frequencies satisfy:

$$\omega_\mu = \omega_0 + D_1\mu + \frac{D_2}{2!}\mu^2 + \frac{D_3}{3!}\mu^3 + \cdots$$
$$= \omega_0 + D_1\mu + D_{int}(\mu), \tag{3}$$

$$D_i = \frac{\partial^i \omega_\mu}{\partial \mu^i}\bigg|_{\mu=0}, \tag{4}$$

where $\omega_0$ is the frequency of the resonant field corresponding to the pump, and μ is the relative mode number counted with $\omega_0$. $D_1/2\pi$ is the FSR of the cavity around the pump frequency $\omega_0$, for FSR varies with mode number influenced by dispersion. $D_2/2\pi$ is the walk-off of resonances from an equidistant resonance grid spaced by $D_1$, which directly and mainly reflects the group velocity dispersion of the mode, satisfying:

$$D_2 = \omega_{+1} - 2\omega_0 + \omega_{-1} = -\beta_2 D_1^2/\beta_1 \tag{5}$$

where $\beta_1$ is the reciprocal of group velocity, and $\beta_2$ is the derivative of group velocity with respect to angular frequency, called group velocity dispersion. $D_2$ and higher-order dispersion $D_i$ ($i \geq 3$) account for the integrated dispersion $D_{int}$. From the cavity transmission spectrum in **Figure 3a**, it can be seen that there are two mode families in the cavity with different FSR, which will induce an avoided-mode crossing at a certain frequency.

Previously, it was thought that in a perfect isotropy resonator without any defect from fabrication, this crossover could be eliminated for cavity eigenmodes are orthogonal and there is no interaction between them. By comparing the shape of the transmission spectra of these two mode families under high power TE pump, we find that resonances of the mode with higher transmittance (the hollow circle) does not show triangular distortion caused by thermal effect. Thus, we judge this mode to be TM mode with orthogonal polarization of pump (TE), which is consistent with previous reports of both TE and TM modes supported in birefringent resonators [6, 9]. The integrated dispersion $D_{int}(\mu)$ of TE mode (solid line) in Figure 3b is fitted by Equation 3 with fourth-order dispersion from a uniform FSR ($D_1/2\pi \approx 905.32\ GHz$) of TE mode. The positive $D_2/2\pi \approx 1.10\ GHz$ show this waveguide possesses an anomalous dispersion, which is essential for a soliton formation. It is clear that an avoided-mode crossing occurs between Pump 2 and its right cavity mode (blue-wavelength side), giving the reason why one sideband of a pair is suppressed in stages II and III of nondegenerate FWM depicted in Figure 2b.

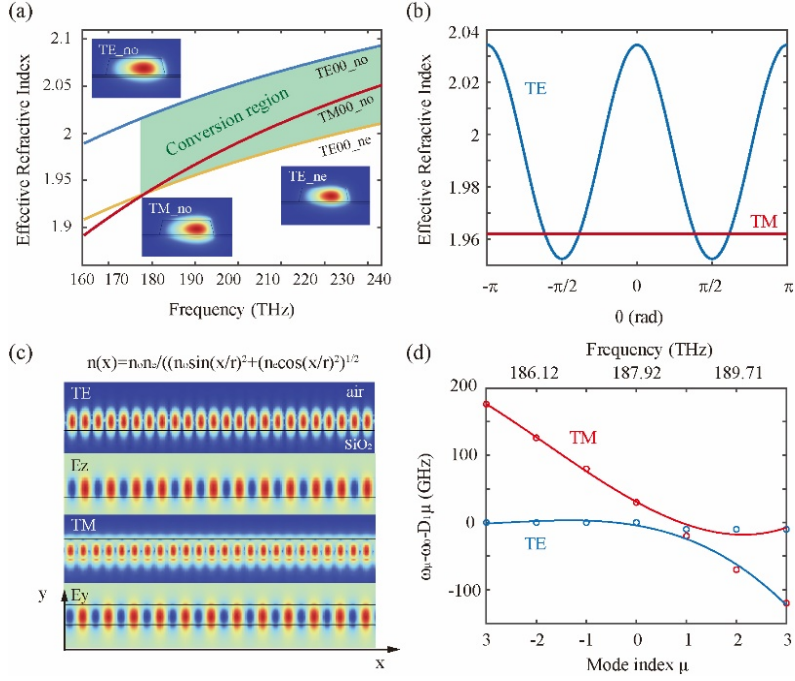

**Figure 4.** Analysis of the polarization conversion of TE and TM modes in X-cut rings. a) The effective refractive index of TE mode propagating along the extraordinary (TE_ne) and ordinary (TE_no) axis and that of TM mode along the ordinary axis (TM_no) in the ring, obtained by COMSOL simulation. Green region shows the range where polarization conversion happened. b) Effective refractive index changes versus angle for TE mode in cavity at 1594 nm pump (blue line), Pump 2, and the index for TM mode (red line). c) 2D modal for simulating eigenmodes of X-cut rings. The microring is equated to a straight waveguide with periodic boundaries and periodic variation of refractive index. The y-axis, from bottom to top, represents substrate (SiO$_2$), LN, air. d) Simulated cavity resonances (eigenmodes of the waveguide), TE (blue) and TM (red), and avoided-mode crossing in the frame of TE mode. The pump is at 187.92 THz. Dispersion is neglected here.

Next, we will analyze the origin of this mode crossing, hypothesis of the coexistence of TE and TM modes in X-cut LN microrings under TE or TM pump. In X-cut LN microcavities, the optical axis Z lies in the plane, as shown in the coordinate system of Figure 1a, so the effective refractive index of TE mode in the cavity is:

$$n_{eff}(\theta) = \frac{n_o n_e}{(n_o^2 sin^2\theta + n_e^2 cos^2\theta)^{\frac{1}{2}}}, \quad (6)$$

within the interval $[n_e, n_o]$. $\theta$ is the angle between the optical axis and the wavevector. $n_o$ and $n_e$ denote the effective refractive indices of ordinary and extraordinary wave in LN crystal respectively, which can be simulated by setting isotropic material properties in COMSOL, as shown in **Figure 4a**. For TM mode, the polarization direction is always parallel to the X axis, so its effective index is the index of ordinary wave, $n_o$. From Figure 4a, it's depicted that the effective refractive index of TM mode lies between $n_o$ and $n_e$ of TE mode after the crossing at 180 THz, the blue region, indicating that TM mode will be bound to have the same effective index as the TE mode at 4 azimuths of the ring, according to Equation 6, where these two orthogonal modes have strong interaction and polarization conversion. For example, the blue line in Figure 4b is the effective index of Pump 2 as a function of angle, while the red line shows that of TM mode. In this simulation, the material definition is isotropy, so that the effective $n_o$ and $n_e$ can be calculated and then substituted into Equation 6 to present the effective index of TE mode. The intersections near $\theta = \pm \pi/2$ suggest that here the polarization is converted, and the TM mode will be stimulated when it resonates in the cavity. As the FSR mismatch of TE and TM mode, there will be an avoided-mode crossing around one certain resonance. Thus, it will suppress the generation of one sideband of FWM and disrupt the formation of cascaded FWM, which will also impose a negative effect on the formation of soliton. Insets in Figure 4a display the electric field distribution of TE and TM modes of the microring in our experiment under isotropic material definition. Although, it has been reported that mode coupling between various orders of TE modes may occur in LN waveguides, the conversion efficiency or crosstalk is about 10 to 20 dB [13, 14]. In our experiment, shown in Figure 3a, there is a broadband conversion with about twofold difference between TE and TM modes, where TM modes have weaker coupling between bus waveguide and ring, so the mode-order conversion can be neglected as well as polarization conversion with TM modes.

In isotropic materials, it is easy to calculate the eigenmodes and mode crossing point by calculating the effective refractive indices of TE and TM modes then substituting them to resonance condition. However, this practice doesn't apply to TE modes in X-cut rings due to birefringence, while it is feasible for TM modes. By anisotropic material definition, it may be possible to simulate in COMSOL by 3D modal, which has not been reported as far as we know. The huge computational amount is the biggest obstacle in this occasion. Therefore, one effective 2D modal is proposed to predict the position of the unavoidable avoided-mode crossing. Here, the microring is regarded as a straight waveguide with periodic boundaries. Periodicity is the circumference of a ring, and the birefringence is replaced by an effective index equation in Figure 4c, a function of x, where $x = \theta r$ denotes the propagation distance of light within the ring. Since the 2D model simplifies the field distribution in z-direction, infinite waveguide width, the length of straight waveguide is different from that of ring. So, we use the equivalent circumference of 26.1 um in our simulation. In order to preserve the influence of substrate (SiO$_2$) and air on the mode distribution as much as possible, we analyze the eigenmodes from the profile as depicted in Figure 4c. In y-axis direction, the effective index is defined as $n_o$, and in z-axis direction, that is similar to Equation 6 but with a $\pi/2$ shift. Figure 4c shows the eigenmodes distribution of TE (187.92 THz) and TM (187.95 THz) and their field components at z and y direction, respectively. The calculated resonances are plotted in Figure 4d without considering dispersion, which agree with the experimental results in Figure 3b well.

## 3. Conclusion

In this paper, we complement the theory of polarization conversion in birefringent waveguides, that in addition to the geometric structure and scattering of annular waveguides, the natural property of the periodic fluctuation of mode effective index with angle in birefringent crystals can cause the conversion between TE and TM modes. This mechanism makes the coexistence of these two orthogonal modes in X-cut LN microrings unavoidable, as well as the avoided-mode crossing between these two mode families due to FSR mismatch. The induced mode crossing will distort the dispersion heavily [29], which suppressed one sideband of the local nondegenerate FWM and the cascaded FWM later, imposing a disruption on the formation of soliton possibly. Then to calculate the mode crossing point and avoid the obstacle from huge computational effort, we proposed a 2D method by stretching the ring to a straight waveguide with periodic boundaries and periodic variation of effective index. Simulated resonances agree reasonably well with the experiment by this method, which is the first trying to our best knowledge. This work paves the way for future theoretical research and applications of polarization conversion in birefringent waveguides, like calculating conversion efficiency by coupled-mode equation and designing a mode converter just by a short-curved waveguide with small footprint. For example, it will provide suggestions for the new design of entirely integrated polarization converters, which can be embed in quantum circuits, simulation on birefringent platform and maybe helpful for the formation of soliton in X-cut LN microrings.

**Supporting Information**

Supporting Information is available from the Wiley Online Library or from the author.

**Acknowledgments**

We thank Weimin Ye, Ning Liu, Jipeng Xu and Xingqiao Chen for valuable discussions during the preparation of this manuscript. This work is supported by National Natural Science Foundation of China (12274462, 11674396), Department of Science and Technology Department of Hunan Province, China (2017RS3039, 2018JJ1033) and Hunan Provincial Innovation Foundation for Postgraduate, China (QL20210006).